\documentclass[10pt,journal,compsoc]{IEEEtran}
\ifCLASSOPTIONcompsoc
  \usepackage[nocompress]{cite}
\else
  \usepackage{cite}
\fi
\ifCLASSINFOpdf
   \usepackage[pdftex]{graphicx}
   \graphicspath{{../pdf/}{../jpeg/}}
   \DeclareGraphicsExtensions{.pdf,.jpeg,.png}
\else
\fi
\ifCLASSOPTIONcompsoc
  \usepackage[caption=false,font=footnotesize,labelfont=sf,textfont=sf]{subfig}
\else
  \usepackage[caption=false,font=footnotesize]{subfig}
\fi
\usepackage[latin1]{inputenc}

\begin{document}
%
\title{Teaching Performance Modeling in the era of millennials}
%
%
%

\author{Vittoria~de~Nitto~Person\'e,~\IEEEmembership{Member,~IEEE,}
\IEEEcompsocitemizethanks
{\IEEEcompsocthanksitem 
V. de Nitto Person\'e was with the Department
of Civil and Computer Engineering, University of Rome Tor Vergata, Italy,
\protect\\
E-mail: denitto@ing.uniroma2.it 
}
\thanks{Manuscript received January 23, 2020; revised.}}

%
%

\markboth{Journal of \LaTeX\ Class Files}%
{de~Nitto~Person\'e
 : Bare Demo of IEEEtran.cls for Computer Society Journals}
%



\IEEEtitleabstractindextext{%
\begin{abstract}
Performance Modeling (PM) teaching started in the early 70's and reached its peak in the 80s. From those years and until today computing systems have deeply changed. Moreover, in the last two decades an economical crisis has involved the educational system, while the new generations show new learning modes. In this time, new literature has developed about learning and teaching. Rarely highlighting the critical issues. Higher learning is changing its role maybe unawares. In this paper, the author starts from a close examination of the state of the art of PM courses in Universities around the world and tries to highlight the main critical issues of teaching nowadays. The paper has not the aim to give answers but  mainly that to open the way to reflections and discussions.

\end{abstract}

}

\maketitle

\IEEEdisplaynontitleabstractindextext

%
\IEEEpeerreviewmaketitle

\IEEEraisesectionheading{\section{Introduction}\label{sec:intro}}

%
%
%
%
\IEEEPARstart{T}{he} roots of Performance Modeling were planted in the 70's.
In June 1959, the first International Conference on Information Processing was held in Paris, under the sponsorship of UNESCO. During that conference, the International Federation for Information Processing (IFIP) was established to meet the need to promote information science and technology, stimulating research, development and cooperation among several countries. Among the others, one important aim was "encouraging education in information processing" \cite{ifip}. In 1972, the Technical Committee TC 7 System Modeling and Optimization was established. The WG 7.3 Computer System Modeling was one of its  working groups. In 1973, the WG 7.3 - International Symposium on Computer Performance Modeling, Measurement, and Evaluation started its activity.

During the same years, the National Bureau of Standards and its Institute for Computer Sciences and Technology started a series of Federal Information Processing Standards (FIPS) Task Groups. In 1971, the FIPS Task Group 10 Computer Component and Systems Performance Evaluation promoted "a self-governing Computer Performance Evaluation User's Group (CPEUG) whose purpose is to disseminate improved techniques in performance evaluation through liaison among vendors and Federal ADPE users, to provide a forum for performance evaluation experiences and proposed applications, and to encourage improvements and standardization in the tools and techniques of computer performance evaluation." \cite{cpeug}. The CPEUG collected people "from many United States Governmental agencies involved in various phases of this field a number of academicians as well as analysts from business and industry working in this area, and this gave rise to the formation within the ACM of SIGME [Special Interest Group in Measurement and Evaluation] which is currently known as SIGMETRICS". In 1974, for the first time the proceedings of 8th meeting of CPEUG was made available as "a major source in the limited literature on computer performance, evaluation and measurement". It is quite interesting what we can read in the preface: "Computer performance, evaluation and measurement is now vital to the designer, the user and the management-owner of a modern computer system.
To some, computer performance, evaluation and measurement is a tool, a marriage of abstract thought and logic combined with the techniques of statistical and quantitative methods. To others, it is a technique with very heavy reliance on modeling and simulation and simultaneously involves features of both classical experimentation and formal analysis. The problem of exact specification is made the more difficult by the recent birth and development of computer performance, evaluation and measurement as a discipline within computer science". 

In ten years most universities activated Performance Modeling (PM) courses and the most are still active today.
From that starting time a long path has been followed. Computing systems have deeply changed. The technological advances have made possible concepts and systems that were surely unimaginable at that time. Hand in hand the curricula had to adapt. 

In this work, a close examination of the current courses in PM over the world is presented and a classification is proposed. 
Starting to make the point about PM teaching, one has to face some general questions. 
There are a lot of recent activities in learning and instruction, but infrequently they point to the general difficulties in which the teaching activity currently is. At least in a part of the world, the educational system is living a critical period. The role of higher education is changing, while the new generation of students has already changed. In this paper, the author focuses on some aspects in order to stimulate reflections and to start discussion with the firm belief that the best way to face change is awareness. A preliminary version of this paper was presented in  \cite{140}.

The paper is organized as follows. Section \ref{sec:pmc} presents the state of the art of Performance Modeling courses around the world. The courses are classified according three main categories, the category contents together with the distribution of courses in the world are presented. Appendix A presents the full list of Performance Modeling courses organized by categories. Section \ref{sec:crisis}  considers the role that Performance Modeling still has today and introduces the issue of the importance of knowledge basic methodologies to face with  new technologies in constant evolution. Section \ref{sec:crisisArg} points to two other critical aspects that heavily influence teaching. Finally, Section \ref{sec:conc}  concludes the paper. 

 

\section{Performance Modeling Teaching}
\label{sec:pmc}
In this section, first the search methodology is briefly explained, then the currently taught courses are identified and classified.

\subsection{Research Methodology}
The search for PM teaching has been carried out by considering the relevant communities. To this aim, the considered groups include the WG 7.3 Computer System Modeling, as it appears in the official IFIP web page (http://www.ifip.org/bulletin/bulltcs/memtc07.htm), and program committees and editorial boards of some specialized conferences and journals. In particular:

The program committees of the following conferences: Sigmetrics 2015, ACM/SPEC International Conference on Performance Engineering 2017, IEEE International Symposium on the Modeling, Analysis, and Simulation of Computer and Telecommunication Systems 2017, the Workshop on MAthematical performance Modeling and Analysis 2017, and the first Workshop on Education and Practice of Performance Engineering 2017,
the Editorial boards of the ACM Transactions on Modeling and Computer Simulation and of the ACM Transactions on Modeling and Performance Evaluation of Computing Systems,
and the Italian group on Quantitative Methods in Informatics, InfQ.

Globally, more than 300 web pages have been consulted. Sometimes the access to degree requirements and program courses is available, but often it is not. In this last case, the reference colleague has been directly contacted by email. 
Just currently taught courses are considered.


\subsubsection{Current Courses Classification}
At the time of this search, about 75 courses are taught, both at the undergraduate and graduate levels. These courses share the modeling approach, that is the definition of a model to evaluate the characteristics and the behaviour of the "system" under study. Courses from Operations Research area, or "pure" Queueing Theory courses, or courses on Simulation in different contexts are excluded from this search.

\begin{figure*}[!t]
\centering
\subfloat[Canada US] {\includegraphics[width=2.5in]{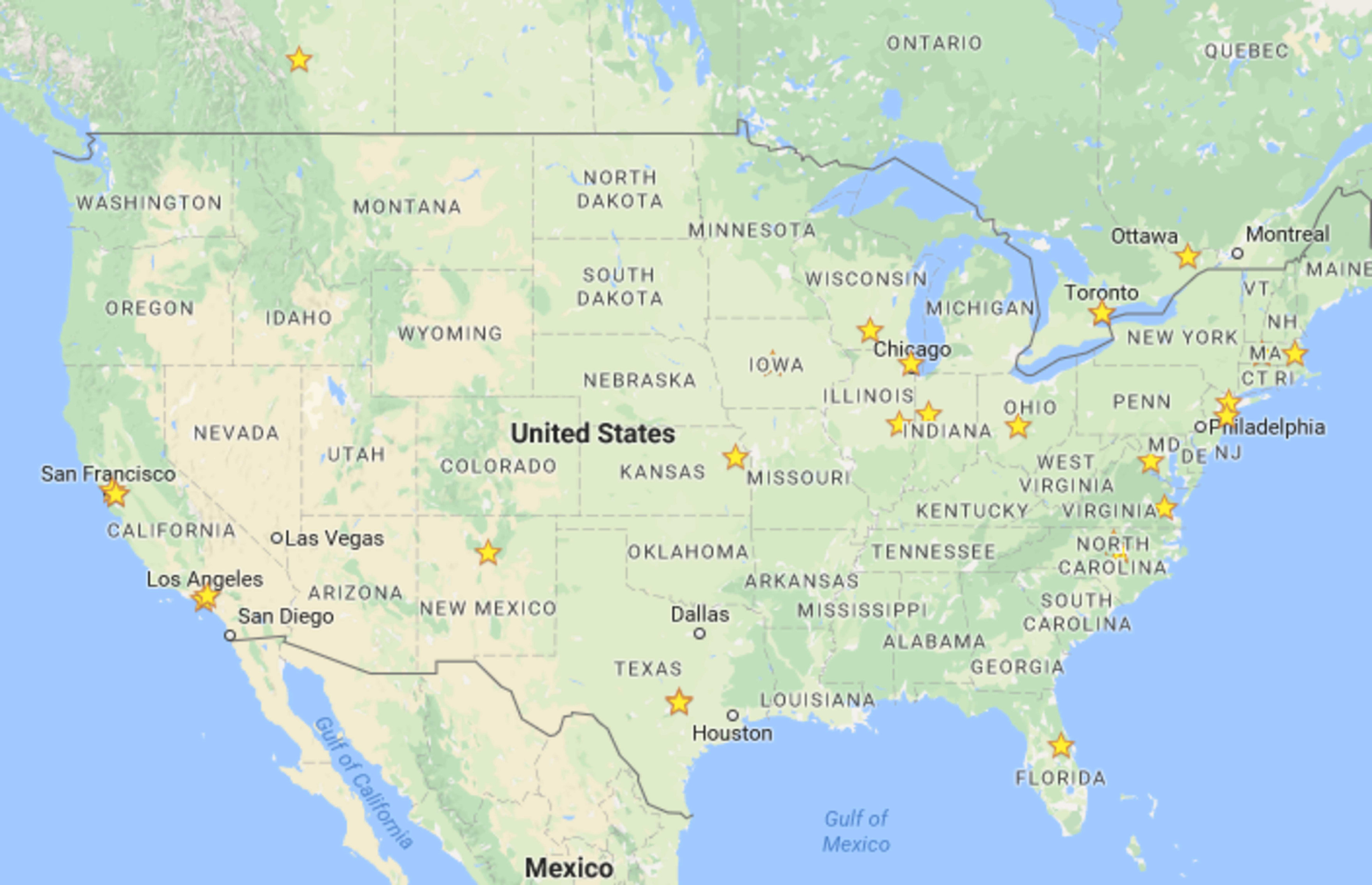} %
\label{CUS map}}
\hfil
\subfloat[Europe] {\includegraphics[width=2.5in]{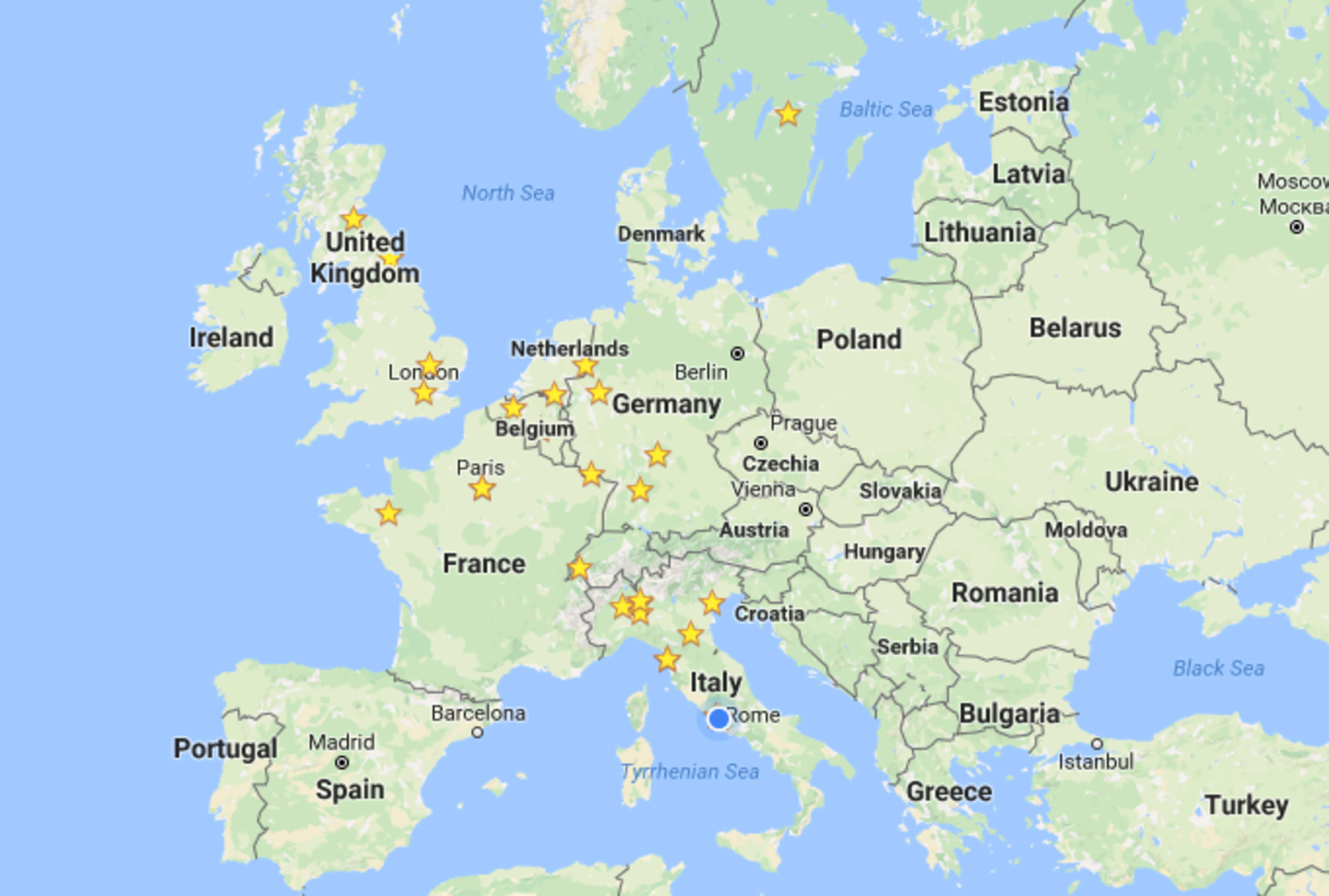}%
 \label{EU map} }
 \hfil
\subfloat[Asia] {\includegraphics[width=2.5in]{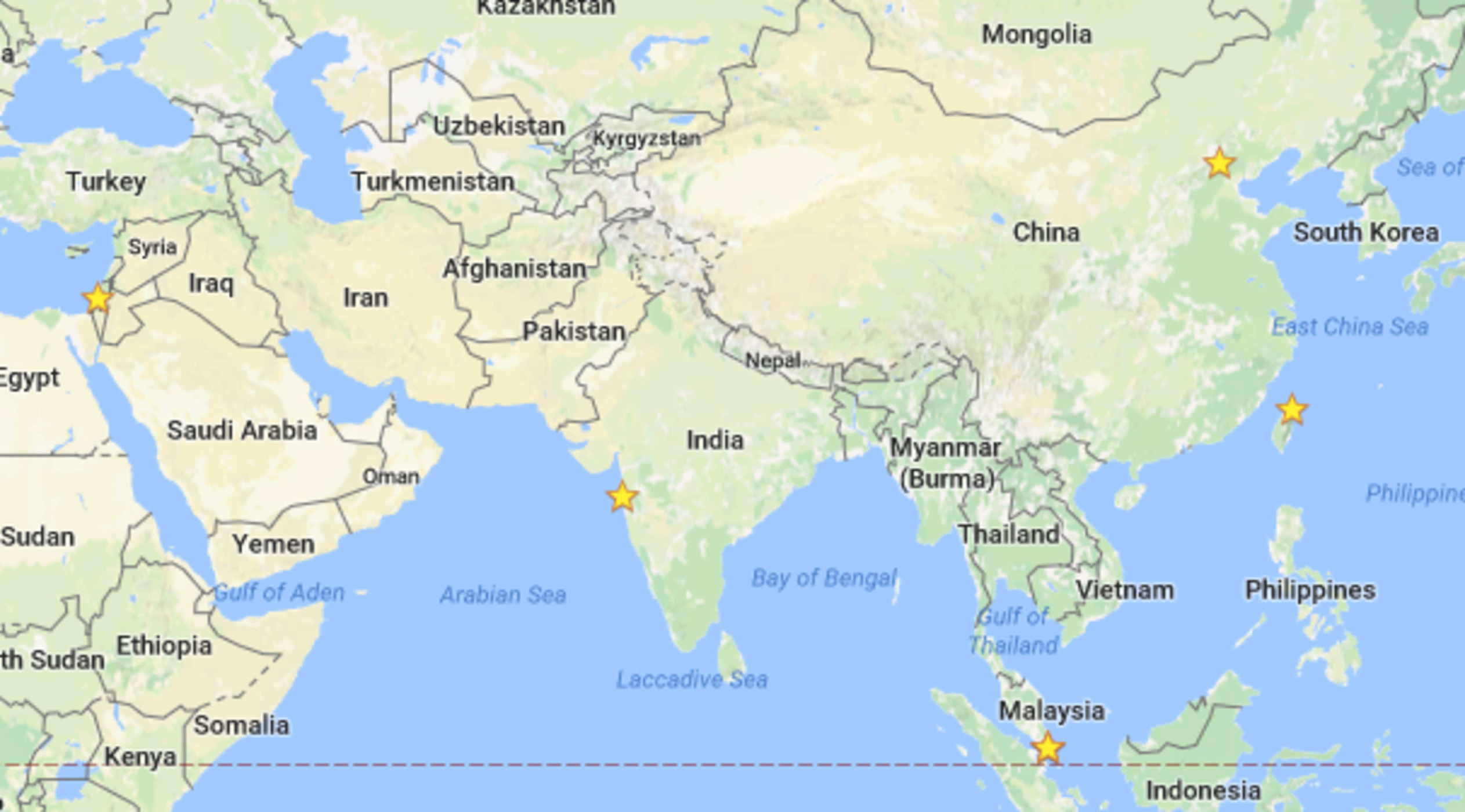}%
 \label{Asia map} }
\caption{PM teaching geographic map}
\label{fig_map}
\end{figure*}

Three different areas are selected: Europe (EU), Canada and United States (Can/US) and Asia. In Fig.\ref{fig_map}, the locations of the PM courses are shown in the three different geographic areas. Note that one star may means more than one course. 

The courses can be partitioned in different groups, so to identify common applications domain and aims. The differentiation is not always so clear and some course could belong to more than one groups. Three groups have been defined: $General~PM$, $PM~for~Communications~Systems$, $PM~for~Software$. Table \ref{tab:table_map} shows the distribution of the PM courses in the geographic areas with respect to the three categories. In Appendix A, the reader can find the complete list of the courses with the course title, the teacher, or more than one if they share the course, and the University where it is taught.

\begin{table}[!t]
\renewcommand{\arraystretch}{1.3}
\caption{PM courses geography }
\label{tab:table_map}
\centering
\begin{tabular}{|c|||c|c|c|}
\hline
& $Gen~PM$ & $PM~for~CS$ & $PM~for~SW$\\
EU & 24 & 12 & 6\\
Can/US & 15 & 7 & 4\\
Asia &  6 & 1 & na\\
total & 45 & 20 & 10\\
\hline
\end{tabular}
\end{table}

It is interesting to identify the common basic methodologies used in the courses. There is a strong common core beyond the different application areas: even at a different extent, all courses use operational laws, queueing systems, and statistics \cite{oper}, \cite{kleinrock}, \cite{allen}. In addition, the used methodologies and the general approach differ with respect to the application perspective and aims.

$General~PM$. In this group the courses are quite homogeneous. Most courses use probability, simulation and stochastic processes. Some courses use a different approach and techniques. 

$PM~for~Communications~Systems$. This group is quite heterogeneous as contents and approaches. Even course titles show greater variability than the previous group.

$PM~for~Software$. To the best of our knowledge, few courses are in this group. As a consequence, deriving general trends make little sense. However, they share a common core and specifically Layered Queueing Networks for modeling complex software systems \cite{lqn}. On the other hand, in USA three companies offer performance engineering courses. The reader can find these in Appendix A.
Table  \ref{tab:table_contents} shows the contents map.

\begin{table}[!t]
\renewcommand{\arraystretch}{1.3}
\caption{Contents map}
\label{tab:table_contents}
\centering
\begin{tabular}{|c|c|c|}
\hline
$Gen~PM$ & $PM~for~CS$ & $PM~for~SW$\\
\multicolumn{3}{|c|}{\textbf{Operational Laws, Queueing Systems, Statistics}}\\
Fluid~models & Combinatorics & LQN \\
Optimization & Control & Petri~nets\\
Petri~nets & Fluid~models & Simulation\\
Probability & Game~Theory & Workload~charact\\
Process~algebras & Graph~Theory & \\
Simulation & Net~Calculus & \\
Stochastic~processes & Optimization & \\
Timed~automata & Probability & \\
Workload~charact & Simulation & \\
 & Stochastic~processes & \\
\hline
\end{tabular}
\end{table}

Finally, we have also noted that some consolidated courses have been canceled or changed to make room for new needs. Some examples follow.


At Ca' Foscari University, Venice - Italy, Simonetta Balsamo taught Performance and Reliability of Computer Systems, a general PM course, until 2015. Then the course was closed 
and she teaches several courses in cloud computing and networks.

At Caltech, USA, Adam Wierman taught Analytic tools for system design, a general PM course, until 2009. During the next six years, Wierman taught Network performance analysis, a PM for CS course. Since 2016, Wierman teaches Networks: Structure \& Economics that can not be categorized as a PM course.

At University of Southern California, USA, in 2015, Konstantinos Psounis changed the title of his course Probabilistic Methods in Computer Systems Modelling in Probability for Electrical and Computer Engineers. The contents are broadly the same.

At the Ohio State University, USA, Cathy H. Xia changed the framework of her course Performance Modeling and Simulation and since 2015 she teaches Simulation for System Analytics and Decision-Making.

At the Chinese University of Hong Kong, China, John C.S. Lui taught Computer System Performance Evaluation until 2009. Now, he teaches several courses in Internet technology and Machine Learning.

\section{Basic knowledge or Specialization}
\label{sec:crisis}
The evolution of computing field is continuous and fast, new technologies and applications enrich the landscape of computer technology. As a consequence, new computing-related disciplines emerge and push to update curricula. 
However, both curriculum guidelines for Computer Engineering and for Computer Science include PM
 \cite{CEcurriculum}, \cite{CScurriculum}. Table \ref{tab:table_CE_CS} shows knowledge areas and relative topics in both guidelines. Modeling, performance and evaluation are important keywords of several knowledge areas. 

\begin{table}[!t]
\renewcommand{\arraystretch}{1.5}
\caption{Knowledge Areas and Topics}
\label{tab:table_CE_CS}
\centering
\begin{tabular}{|c|c|}
\hline
$Knowledge~Area$ & $Topics$\\
\multicolumn{2}{|c|}{\textbf{Computer Engineering \cite{CEcurriculum}}}\\
Comp.~Arch.~and~Organization & Measuring~perform. \\
Computer~Networks & Performance~evaluation\\
Syst.~and~Project~Eng. & Syst.~arch.~design~and~eval.\\
Syst.~Resource~Management & System~perform.~eval.\\
\multicolumn{2}{|c|}{\textbf{Computer Science \cite{CScurriculum}}}\\
Operating Systems & System Perform. Eval.\\
Parallel and Distrib. Comp. & Parallel Perform.\\
Systems Fundam. & Evaluation\\
 & Resource Alloc. \& Scheduling\\
 & Quantitative Evaluation \\
\hline
\end{tabular}
\end{table}

Furthermore, it is also interesting to note a general trend towards specialization. In  the training system, this could lead to a reduction in basic knowledge. So, if on one hand there is the need to update curricula, on the other hand there is the worry of losing knowledge in general principles.


An interesting point of view is offered in \cite{denning}. Denning and Martell analyze computing with a different perspective: computing is a science governed by fundamental principles that span all technologies. Their book has a monumental aim: identify the principles behind computing, principles that are not changed during all the history and that probably will not change for a long time. They divide the great principles into six categories that
 "are like windows of a hexagonal kiosk. Each window see the inside space in a distinctive way; but the same thing can be seen in more than one window". All computing domains build over these common principles. One of this categories is Evaluation and the authors have strongly advocated that performance modeling and engineering are fundamental parts of computer science.

Concern about the effects of losing basic methodologies begins to become tangible. In the framework of modeling, the research community itself observed shortfalls in the use of evaluation methodologies even in papers published in premiere journals and conferences. For example, in telecommunication networks, the lack of rigor and formal approach in simulation studies was highlighted in  \cite{pawlikowski, kurkowski}. The availability of user-friendly simulation packages "...has led to a belief that simulation is mainly an exercise in computer programming" \cite{pawlikowski}, while a valid simulation study requires ability to abstraction, intuition and deep knowledge of probability and statistics. This supports the idea that limiting teaching space of general methodologies can be a dangerous pitfall.

However, other issues affect the educational system nowadays and we introduce the most important ones in next Section.

\section{Economy, Profit and New Generation}
\label{sec:crisisArg}
In today's world, it is easy to observe a trend towards specialization. As introduced in the previous Section, in  the training system this could happen at the expense of basic knowledge.  Moreover, the educational system suffers the consequences of an economical crisis. These two aspects are somewhat correlated. In the following, some points are introduced to open the way to a deeper reflection.

A first important issue has to be considered. Nowadays, there is an economical crisis. One of the main victims has been the educational system, both at low and high levels. Starting from the primary school \cite{leachman} to the university, the educational system has been severely reduced. In southern Europe, the economic crisis is a reality that reflects on education. In Italy, cuts in funds have reached 20\% \cite{demartin}. In US, cuts are differentiated per states but at similar percentages \cite{mitchell}. 

One of the effects is a continuous monitoring activity to identify courses that appear do not be productive. Sometimes, this is evaluated in terms of the number of exams per year. This again could become a penalty for basic courses. However, the idea of $productivity$ is merely related to a job. The opinions that higher education is mainly training of workers or that knowledge must be immediately useful for the economy, are questionable \cite{demartin}.

On the other side, nowadays we observe a hyper-specialization of work, with demands that become increasingly sectorial. This is partially due to the technological revolution, but this is also a side-effect of crisis: industry cannot spend in training and ask universities to produce specialized workers, according to their current needs.
To what extent is the academic world aware of its changed role? 

All this is bringing a social transformation: on one hand it is possible to observe a general decreasing of educational level, and on the other hand we are witnessing an excess of specialization. In our daily life, we often face with people that acknowledge their inability to solve problems, a frequent answer is "this is not my job". In particular, it is easy to observe a general lack of ability to face with unexpected situations. It is possible to speculate that this is the result of the combined effect of the decreasing educational level on one side and of the hyper-specialization on the other side. From this latter point of view, the excess of training in use of technology rather than the knowledge of the underlying principles brings in itself this risk. The use of computer-aided tools (CAD) in infrastructure projects is a good example of this risk. It is worth noting that CAD tools, and the technology in general, are extremely useful instruments, but their appropriate use requires anyway deep knowledge of processes or methodologies that, for example, bring to the project result. Unfortunately, some catastrophic events begin to highlight some lack of expertise.

If on one hand the university is pushed to become a utilitarian organization and on the other hand the technological tools are substituting knowledge, how does the education community want to answer to this challenge? Does it want to recuperate the thinking spaces? Or does it want to completely abdicate to the economic/technological domain? In other words, a reflection is needed on the role of higher education, the role of University.

Satish K. Tripathi was one of the first members of WG 7.3 Computer System Modeling. Now, he is the President of Buffalo University and his viewpoint is \cite{tripathi}: "The aim of higher education is not merely to prepare students for jobs. It is to prepare them to lead, innovate, and contribute meaningfully to the world around them."

Drew Gilpin Faust is the President of Harvard University. Her point of view is extremely interesting \cite{drew}: "Higher learning can offer individuals and societies a depth and breadth of vision absent from the inevitably myopic present. Human beings need meaning, understanding and perspective as well as jobs. The question should not be whether we can afford to believe in such purposes in these times, but whether we can afford not to".

There are no doubts: critical thinking and ability to face changing challenges are the real skill that a high degree instruction should have among its aims. But how to face with the new generation and their new minds? 
This is the second big issue that teachers have to consider. As mentioned in Section \ref{sec:intro}, Computer Systems have deeply changed, the technological advances have made possible concepts and systems that were surely unimaginable even some decades ago. At a slow pace, the world has changed too: the way we work, the way we interact with each other, the way children grow is completely different from the way we ourselves grew half a century ago. This has changed the way the mind works and learning process itself has also quite changed. Change leads to difficult times, but it is positive. However, awareness is essential to go through change. There are no precise recipes but the firm belief that teaching has to consider this and has to find new ways to be effective.

As highlighted above, the new generation is completely different from students of just a decade ago. Michel Serres loves this young generation \cite{serres}: they have huge amount of information handy, and they have access to any notion, concept, method, anywhere, anytime. The access to information is completely different from accessing organized information as we had before internet, and another important aspect is that this access is without mediators. Putting off at the deep thought of Serres and at his great expressive capacity, this is not a negative, but is a new potential and open to new learning modes. Therefore, our lectures must consider it. The interest of the new generation cannot be captured by teaching what they can easily catch by themselves. But Serres advocates: Information is not Knowledge. 

A hard task is once again on our side: the excess of Information need to be transformed in Knowledge, Thumbelina (Serres's young heroine in today's world) needs of our help to do this. And as the new thumbelinas are extremely intelligent, all surface stratagems aimed to catch the attention using charming keywords or appealing communication tools are destined to fail. There are no guidelines or suggestions for a succesful teaching, except for keeping in our mind who we have in front. By using the words of Serres, "Before to teach someone something, you must at least know him".

\section{Conclusion}
\label{sec:conc}
In this paper, first the PM teaching has been presented as it is currently taught. A geographic map and a classification based on the contents have been presented. By considering some critical issues met during the last decade, general issues about teaching and the role of higher education have been highlighted. It is worth noting that recently there have been a lot of activities (e.g. meetings and specialized journals) devoted to teaching and learning, but very few of them have explicitly addressed the points of criticality considered in this paper.

This paper pushes for uncover the criticality. The economical crisis that is reducing the educational system and the influence of industry interests are pushing University to become utilitarian organization. Moreover, the change of new generation asks for new teaching way. If the education community does not want to abdicate to the economic/technological domain and if it wants to defend its leading role in knowledge and critical thinking, it need to be fully aware of the surroundings difficulties. The aim of this paper is to open the way to reflections and discussions.


%

\appendices
\section{The PM courses}
In the following, the courses are grouped according to the classification in Section \ref{sec:pmc} and are listed alphabetically.
The information are organized in the following order: course title, teachers and university.

The author apologize for all colleagues that have been involuntarily skipped.

\subsection{General PM courses}
Analysis of production systems, Ivo Adan, Eindhoven University of Technology, The Netherlands

Analytical Modeling of Computing Systems, Varsha Apte, Indian Institute of Technology, India

Analytical Performance Modeling, Kristy Gardner, Amherst College, USA

Analytical Performance Modeling \& Design of Computer Systems, Mor Harchol-Balter, Osman Yagan, Carnegie Mellon, USA

Analytical Performance Modelling for Computer Systems, Yong C. Tay, National University of Singapore, Singapore

Capacity Planning, Bruno Ciciani, "Sapienza" University of Rome, Italy

Computer System Analysis, Marco Scarpa, University of Messina, Italy

Computer Systems Analysis, David M. Nicol, William Sanders, University of Illinois, Urbana-Champaign, USA

Computer Systems and Performance Evaluation, Derek Eager, University of Saskatchewan, Canada

Computer System Performance Evaluation, Daniel A. Menasce, George Mason University, USA

Computer Systems Performance Analysis, Teo Yong Meng, National University of Singapore, Singapore

Computer Systems Performance Evaluation, Giuseppe Serazzi, Marco Gribaudo, Politecnico di Milano, Italy

Fundamentals of Computing Systems, Azer Bestavros, Renato Mancuso, Boston University, USA

Introduction to Computer Performance Modeling, Harry G. Perros, Do Y. Eun, North Carolina state University, USA




Methodologies For Discrete-Event Modeling And Simulation, Gabriel Wainer, Carleton University, Canada

Modeling and Performance Evaluation, Vishal Misra, Columbia University, USA

Modeling and Simulation, Peter Buchholz, University of Dortmund, Germany

Modelli e linguaggi di simulazione, Giuseppe Iazeolla, University of Rome Tor Vergata, Italy


Performance Analysis of Computer Systems and Networks, Varsha Apte, Indian Institute of Technology, India

Performance Evaluation, Anne Buijsrogge, Pieter-Tjerk De Boer, University of Twente, The Netherlands

Performance Evaluation, Jean-Yves Le Boudec, Ecole Politechnique Federale de Lausanna, Switzerland


Performance Evaluation of Computer and Communication Systems, Cheng-Fu Chou, National Taiwan University, Taiwan

Performance Evaluation of Computer Systems, Marco Gribaudo, Politecnico di Milano, Como Campus, Italy 

Performance Evaluation of Computer Systems and Networks, Giovanni Stea, University of Pisa, Italy



Performance Modeling of Computer Systems and Networks, Vittoria de Nitto Person\'e, University of Rome Tor Vergata, Italy

Performance Modelling, Jane Hillston, University of Edinburgh, Scotland

Probabilistic Models and Data Analysis, Verena Wolf, University of Saarbrucken, Germany 

Probability for Electrical and Computer Engineers, Konstantinos Psounis, University of Southern California, USA

Quantitative Methods and Experimental Design in CS, Daniel A. Menasce, George Mason University, USA

Quantitative Evaluation of Embedded Systems, Anne Remke, Arnd Hartmanns, University of Twente, The Netherlands

Quantitative Evaluation of Embedded Systems, Pieter Cuijpers, Eindhoven University of Technology, The Netherlands
 
Queueing Analysis and Simulation, Dieter Fiems, University of Ghent, Belgium

Queueing systems, Sem Borst, Jacques Resing, Eindhoven University of Technology, The Netherlands

Queueing Theory, Uri Yechiali, Tel Aviv University, Israel 

Simulation, Evgenia Smirni, College of William \& Mary, Virginia, USA



Simulation and Modeling, Changcheng Huang, Ioannis Lambadaris, Carleton University, Canada

Simulation and Modelling, Giuliano Casale, Tony Field, Imperial College, United Kingdom


Simulazione di sistemi, Lorenzo Donatiello, University of Bologna, Italy


Stochastic decision theory, Bert Zwart,  Eindhoven University of Technology, The Netherlands

Stochastic performance modelling, Stella Kapodistria, Eindhoven University of Technology, The Netherlands

Stochastic Processes,  Ioannis Lambadaris, Amir Banihashemi, Carleton University, Canada

System Evaluation, Nigel Thomas, Newcastle University, United Kingdom

Systems Modelling and Analysis, Peter Marbach, University of Toronto, Canada


Valutazione delle Prestazioni, Giuliana Franceschinis, University of Piemonte Orientale, Italy 

Valutazione delle Prestazioni: Simulazione e Modelli, Gianfranco Balbo, Rossano Gaeta, University of Torino, Italy

\subsection{PM for Communications Systems}


Advanced Topics in Computer Networks, Konstantinos Psounis, University of Southern California, USA

Complex networks: theory and applications, Emilio Leonardi, Politecnico di Torino, Italy

Computer Network Performance, Novella Bartolini, Universita "Sapienza" di  Roma




Enterprise Digital Infrastructure, Maria Carla Calzarossa, University of Pavia, Italy

Introduction to Communications Networks, Victor S. Frost, University of Kansas, USA

Introduction to Computer Networks, Peter Marbach, University of Toronto, Canada

Network Analysis and Simulation, Michele Zorzi, Universita di Padova, Italy

Network Modeling, Michele Zorzi, Universita di Padova, Italy

Network modelling and simulation, Marco Ajmone Marsan, Politecnico di Torino, Italy


Performance Evaluation using Queueing Networks, Gerardo Rubino, Bruno Tuffin, University of Rennes, France

Performance Modelling and Simulation, Paul J. Kuhn, Andreas Kirstadter, University of Stuttgart, Germany

Performance Modeling for Mobile Communications Networking Systems, Phone Lin, National Taiwan University, Taiwan

Performance Modelling of Computer Communication Networks, Nicolo Michelusi, Purdue University, USA

Performance of Networked Systems, Rob van der Mei, Vrije Universiteit Amsterdam, The Netherlands

Probability \& Stochastic Processes II, Francois Baccelli, University of Texas at Austin, USA

Random Processes in Communication and Control I, Randall Berry, Northestern University, USA

Simulation and Network Performance, Andrea Marin, Ca Foscari, University of Venice, Italy

Stochastic networks, Frank Kelly, University of Cambridge, United Kingdom

Stochastic networks, Johan S.H. van Leeuwaarden, Sem Borst, Eindhoven University of Technology, The 	Netherlands

The Art \& Science of Quantitative Reasoning, Azer Bestavros, Boston University, USA

\subsection{PM for Software}
Design of High Performance Software, Dorina Petriu, Shikharesh Majumdar, Greg Franks, Carleton University, Canada


Modeling and Measurement of Software Performance, Diwakar Krishnamurthy, University of Calgary, Canada

Performance Engineering, Dorina Petriu, Shikharesh Majumdar, Greg Franks, Carleton University, Canada

Performance Engineering, Giuliano Casale, Holger Pirk, Imperial College, United Kingdom

Performance Engineering, Mark Friedman, University of Washington, USA

Performance Evaluation of Computer Systems, Petr Tuma, Vojtech Horky, Charles University, Czech Republic



Software performance and scalability, Andrea Marin, Ca Foscari, University of Venice, Italy

Software Performance Engineering, Steffen Zschaler, Malcolm Lees, King's College London, United Kingdom

Software Performance Evaluation, Diwakar Krishnamurthy, University of Calgary, Canada  

Software Quality Engineering, Vittorio Cortellessa, Universita dell'Aquila, Italy

\subsection{PM for Business}
In the following, three companies that offer performance engineering courses are listed. The courses offered follow the information about the company.

Performance Dinamics Company, Neil J. Gunther, Castro Valley, California, USA:

Guerrilla Capacity and Performance



Performance Engineering Services, L\&S Computer Technology, Inc., Connie U. Smith, Santa Fe, New Mexico, USA:

Performance Solutions: Solving Performance Problems Quickly and Effectively

Software Performance Engineering: Methods and Quantitative Techniques for Proactively Managing Software Performance

Performance Engineering Model Bootcamp: Practical Techniques for Modeling Your Systems

Software Performance and Scalability Consulting LLC, Andre B. Bondi, Red Bank, New Jersey, USA:

Foundations of Performance Engineering

Performance Requirements Engineering and Practice for Product Managers


\ifCLASSOPTIONcompsoc
  \section*{Acknowledgments}
\else
  \section*{Acknowledgment}
\fi

The author would like to thank all colleagues that kindly answered with information about their teaching activity.
In particular, the author would like to thank Evgenia Smirni and Giuseppe Serazzi for helpful discussions and suggestions that really improved the quality of the paper.

\ifCLASSOPTIONcaptionsoff
  \newpage
\fi



%

%

\begin{IEEEbiographynophoto}{V. de Nitto Person\'e}
 is an Associate Professor in Computer Engineering
at the University of Rome Tor Vergata. In 1984, she received the Laurea degree in Computer Science summa cum laude from the University of Pisa, Italy. Her research activity lies in the field of modeling and performance evaluation  of computer/communication systems. 
Main applications include finite capacity and blocking systems, parallel systems, wireless systems and networks, web systems, cloud systems, scheduling and admission control. She has published monographs and papers in international journals, books and conferences. Over 30 years of experience in Performance Modeling teaching and in motherhood nurtured her strong love for Education.
\end{IEEEbiographynophoto}







\end{document}